\documentclass[pra,twocolumn,amsfonts,amsmath,crop,tikz]{revtex4-2}

\usepackage[colorlinks,citecolor=red,urlcolor=blue,bookmarks=false,hypertexnames=true]{hyperref} 
\usepackage[T1]{fontenc}
\usepackage{graphicx}
\usepackage{babel}
\usepackage{braket}
\usepackage{qcircuit}

\begin{document}

\title{Expanding variational quantum eigensolvers to larger systems by dividing the calculations between classical and quantum hardware}

\author{John P. T. Stenger}
\affiliation{NRC Postdoctoral Associate, U.S. Naval Research Laboratory, Washington, DC 20375, United States}
\author{Daniel Gunlycke}
\affiliation{U.S. Naval Research Laboratory, Washington, DC 20375, United States}
\author{C. Stephen Hellberg}
\affiliation{U.S. Naval Research Laboratory, Washington, DC 20375, United States}

\begin{abstract}
We present a hybrid classical/quantum algorithm for efficiently solving the eigenvalue problem of many-particle Hamiltonians on quantum computers with limited resources by 
splitting the workload between classical and quantum processors.  
This algorithm reduces the needed number of qubits at the expense of an increased number of quantum evaluations.  We demonstrate the method for the Hubbard model and show how the conservation of the $z$-component of the total spin allows the spin-up and spin-down configurations to be computed on classical and quantum hardware, respectively.  Other symmetries can be exploited in a similar manner.
\end{abstract}

\maketitle

\section{Introduction}

When quantum computers meet their full potential, they will be able to drastically out perform classical computers for a certain class of problems~\cite{Feynman1982,Lloyd1996,Boixo2018,Shor1997,Cirac2012,Georgescu2014}.  One example is determining the ground state of many-particle interacting quantum systems~\cite{tarruell2019,Salfi2016,Hensgens2017,Singha2011,smith2019,Bakr2009,Schreiber2015,Choi2016,Bordia2017,Mitra2017}.  
With the current state of Noisy Intermediate Scale Quantum (NISQ) devices, it is impractical to have only algorithms that put the burden of computation either totally on the quantum computer or totally on the classical computer.  
Algorithms that split the workload between the classical and quantum devices have the potential to exceed the performance 
of either device alone.  In this way, algorithms can evolve alongside the hardware, shifting more of the work onto the quantum machines as their performance become more robust.

The variational quantum eigensolver (VQE) is anticipated to be one the first algorithms that will show quantum superiority~\cite{Peruzzo2014,McClean2016,Kandala2017}.  In the VQE algorithm, the quantum state is prepared by applying a set of parameterized quantum gates.  Expectation values are calculated on the quantum device using the quantum state and then the parameters are optimized on a classical computer using the output from the quantum computer.  VQE is already a nice example of an algorithm that splits the work between the classical and quantum devices.  However, the limitation of current quantum hardware allows for only very short quantum circuits.
Recent works have proposed shifting more of the algorithm to the classical computer by approximating parts of the simulated system using tensor networks~\cite{Yuan2021,Peng2020}, and by approximating the whole system using classical shadows~\cite{Haung2020,Paini2021}.  We propose algorithms which reduce the number of qubits by shifting the calculation of part of the simulated system onto the classical computer without making approximations.    

In this work, we propose hybrid quantum-classical algorithms to compute the ground state of many body, fermionic Hamiltonians.  First, we present a generalized approach for splitting a quantum many-particle system into two sections and a VQE type algorithm which can deal with one section classically and the other quantum mechanically. In this generalized approach, there are no restrictions on how the system is split.  If the available classical resources have been maxed out, then each additional one-particle state would have to be included in the set treated by quantum hardware.  Our algorithm scales in this case polynomially with an increasing number of one-particle states.  On the other hand, if the available quantum resources have been maxed out, then each additional one-particle state would have to be included in the set treated exclusively by classical hardware.  Our algorithm would then scale exponentially with an increasing number of one-particle states.  Next, we focus on the Hubbard model and split the system into spin-up and spin-down sectors.  The spin up sector is solved on the classical computer while the spin-down sector is solved on the quantum computer.  We will describe three different approaches to solving this spin-split model:  In the first approach, the number of spin-up electrons is fixed while the spin-down space includes all possible electron numbers; in the second, we fix the parity of the spin-down sector; in the third, the total number of electrons in both spin sectors are fixed.  Solving the spin-up sector on the classical computer cuts the required number of qubits in half. Putting restrictions on the spin-down sector can further reduce the number of qubits~\cite{Fischer2019} but can also increase the classical workload.

\section{General Method}
\label{GHSFBV}

The interactions in a physical system of identical fermions can generally be described by the Hamiltonian
\begin{equation}
    \hat{H} = \sum_{\mu\mu'}t_{\mu\mu'}c^{\dagger}_\mu c_{\mu'} + \sum_{\mu\mu'\nu\nu'}v_{\mu\nu\mu'\nu'}c^{\dagger}_\mu c^{\dagger}_\nu c_{\nu'}c_{\mu'},
\end{equation}
where $c^{\dagger}_\mu$ and $c_\mu$ are fermionic creation and annihilation operators acting on the one-particle index $\mu$ in the set $\boldsymbol{\mathrm M}$ and $t_{\mu\mu'}$ and $v_{\mu\nu\mu'\nu'}$ are coefficients specifying the one- and two-particle interactions, respectively.  See the appendix for expressions that can be used to compute the coefficients. 

\subsection{System Splitting}
\label{GMSS}
Our goal is to separate the computational problem into parts such that one part is performed classically and the other part is performed using a modified VQE approach.  To that end, let us begin by defining two subsystems $A$ and $B$ and splitting the one-particle index set $\boldsymbol{\mathrm M}$ into two sets $\boldsymbol{\mathrm M}_A$ and $\boldsymbol{\mathrm M}_B$.  Next, we define for each $\Upsilon\in\{A,B\}$, the subsystem creation operator 
\begin{equation}
	C^{\dagger}_{\vec{n}_\Upsilon} = \prod_{\mu\in\boldsymbol{\mathrm M}_\Upsilon}\big(c^{\dagger}_\mu\big)^{n_\mu},
\end{equation}
where $\vec n_\Upsilon=(n_\mu)_{\mu\in\boldsymbol{\mathrm M}_\Upsilon}$ are particle configurations comprising particle occupation numbers $n_\mu\in\{0,1\}$.  With these operators, we can introduce the Fock states
\begin{equation}
    \ket{\vec{n}} = C^{\dagger}_{\vec{n}_A}C^{\dagger}_{\vec{n}_B}\ket{0},
\end{equation}
where $\ket{0}$ is the vacuum state, which forms a basis for the antisymmetric Fock space $\mathcal F$ for our system.  Rather than mapping the entire $\mathcal F$ to the Hilbert space of the quantum register, however, we only plan to map the antisymmetric Fock space $\mathcal F_B$ for subsystem $B$.  A standard basis for the latter $\mathcal F_B$ can be formed by the Fock states
\begin{equation}
	\ket{\vec{n}_B} = C^{\dagger}_{\vec{n}_B}\ket{0},
\end{equation}
for this subsystem.  For any given particle configuration $\vec n_A$, we can then expand an arbitrary state in $\mathcal F_B$ as
\begin{equation}
    \ket{\Psi_{\vec{n}_A}} = \sum_{\vec{n}_B}\beta_{\vec{n}_A\vec{n}_B}\ket{\vec{n}_B},
\end{equation}
where the coefficients $\beta_{\vec{n}_A\vec{n}_B}$ are normalized such that $\braket{\Psi_{\vec{n}_A}|\Psi_{\vec{n}_A}}=1$.  An arbitrary state in $\mathcal F$ can then be expressed as the linear combination
\begin{align}
	\ket{\Psi} &= \sum_{\vec{n}_A}\alpha_{\vec{n}_A}C^{\dagger}_{\vec{n}_A}\ket{\Psi_{\vec{n}_A}}\nonumber\\
	&= \sum_{\vec{n}_A\vec{n}_B}\alpha_{\vec{n}_A}\beta_{\vec{n}_A\vec{n}_B}C^{\dagger}_{\vec{n}_A}\ket{\vec{n}_B}
\end{align} 
with the normalization $\braket{\Psi|\Psi}=1$.  This form allows us to find any particular state $\ket{\Psi}$, e.g., the ground state of the system, by determining $\alpha_{\vec{n}_A}$ classically and $\beta_{\vec{n}_A\vec{n}_B}$ using our modified VQE approach described below.

VQE requires the calculation of expectation values of $\hat H$ for various states in $\mathcal F$.  The expectation value for the arbitrary state $\ket{\Psi}$ is
\begin{equation}
	\braket{\Psi|\hat H|\Psi}=\sum_{\vec{n}_A\vec{n}_A^{\prime}}\alpha^{*}_{\vec{n}_A^{\prime}}\alpha_{\vec{n}_A}\braket{\Psi_{\vec{n}_A^{\prime}}|C_{\vec{n}_A^{\prime}}\hat HC^{\dagger}_{\vec{n}_A}|\Psi_{\vec{n}_A}}.
	\label{eq:expectationH}
\end{equation}
To evaluate the expectation values on the right-hand side of this equation, we start by grouping the terms in the Hamiltonian
\begin{equation}
    \hat{H} = \hat{H}_A + \hat{H}_B + \hat{H}_{AB}^{(t)} + \hat{H}_{AB}^{(v1)} + \hat{H}_{AB}^{(v2)} + \hat{H}_{AB}^{(v3)}
\end{equation}
such that $\hat{H}_A$ and $\hat{H}_B$ describe interactions exclusively in subsystems $A$ and $B$, respectively, $\hat{H}_{AB}^{(t)}$ describe one-particle interactions involving both subsystems, and $\hat{H}_{AB}^{(v1)}$, $\hat{H}_{AB}^{(v2)}$, and $\hat{H}_{AB}^{(v3)}$ describes two-particle interactions with one, two, and three fermionic operators acting on subsystem $B$, respectively.  The contributions to the expectation value from $\hat{H}_A$ and $\hat{H}_B$, defined by
\begin{equation}
    \hat{H}_{\Upsilon} = \sum_{\mu\mu'\in\boldsymbol{\mathrm M}_\Upsilon}t_{\mu\mu'}c^{\dagger}_\mu c_{\mu'}  + \sum_{\mu\mu'\nu\nu'\in\boldsymbol{\mathrm M}_\Upsilon}v_{\mu\nu\mu'\nu'}c^{\dagger}_\mu c^{\dagger}_\nu c_{\nu'}c_{\mu'},
\end{equation}
can be obtained respectively from
\begin{equation}
	\braket{\Psi_{\vec{n}_A^{\prime}}|C_{\vec{n}_A^{\prime}}\hat H_AC^{\dagger}_{\vec{n}_A}|\Psi_{\vec{n}_A}}=\braket{\vec{n}_A^{\prime}|\hat H_A|\vec{n}_A}\braket{\Psi_{\vec{n}_A^{\prime}}|\Psi_{\vec{n}_A}},
\end{equation}
where $\ket{\vec{n}_A}=C^{\dagger}_{\vec{n}_A}\ket{0}$, and
\begin{equation}
	\braket{\Psi_{\vec{n}_A^{\prime}}|C_{\vec{n}_A^{\prime}}\hat H_BC^{\dagger}_{\vec{n}_A}|\Psi_{\vec{n}_A}}=\delta_{{\vec{n}_A}\vec{n}_A^{\prime}}\braket{\Psi_{\vec{n}_A^{\prime}}|\hat H_B|\Psi_{\vec{n}_A}},
\end{equation}
where $\delta_{{\vec{n}_A}\vec{n}_A^{\prime}}$ is the Kronecker delta.  Similarly, the contributions from terms involving interactions in both subspaces can be obtained from
\begin{widetext}
\begin{align}
	\braket{\Psi_{\vec{n}_A^{\prime}}|C_{\vec{n}_A^{\prime}}\hat{H}_{AB}^{(t)}C^{\dagger}_{\vec{n}_A}|\Psi_{\vec{n}_A}}=\sum_{\mu\in\boldsymbol{\mathrm M}_A}\sum_{\mu'\in\boldsymbol{\mathrm M}_B}(-1)^{N_A}\Big[t_{\mu\mu'}\braket{\vec{n}_A^{\prime}|c^{\dagger}_\mu|\vec{n}_A}\braket{\Psi_{\vec{n}_A^{\prime}}|c_{\mu'}|\Psi_{\vec{n}_A}}-t_{\mu'\mu}\braket{\vec{n}_A^{\prime}|c_\mu|\vec{n}_A}\braket{\Psi_{\vec{n}_A^{\prime}}|c^{\dagger}_{\mu'}|\Psi_{\vec{n}_A}}\Big]
\end{align}
where $N_A$ is the number of particles $\sum_\mu n_\mu$ in the configuration $\vec{n}_A$, and
\begin{align}
	\braket{\Psi_{\vec{n}_A^{\prime}}|C_{\vec{n}_A^{\prime}}\hat{H}_{AB}^{(v1)}C^{\dagger}_{\vec{n}_A}|\Psi_{\vec{n}_A}}=&\sum_{\mu\mu'\nu\in\boldsymbol{\mathrm M}_A}\sum_{\nu'\in\boldsymbol{\mathrm M}_B}(-1)^{N_A}\Big[\big(v_{\mu\nu\nu'\mu'}-v_{\mu\nu\mu'\nu'}\big)\braket{\vec{n}_A^{\prime}|c^{\dagger}_\mu c^{\dagger}_\nu c_{\mu'}|\vec{n}_A}\braket{\Psi_{\vec{n}_A^{\prime}}|c_{\nu'}|\Psi_{\vec{n}_A}}\nonumber\\
    &+\big(v_{\mu\nu'\mu'\nu}-v_{\nu'\mu\mu'\nu}\big)\braket{\vec{n}_A^{\prime}|c^{\dagger}_{\mu}c_{\nu}c_{\mu'}|\vec{n}_A}\braket{\Psi_{\vec{n}_A^{\prime}}|c^{\dagger}_{\nu'}|\Psi_{\vec{n}_A}}\Big],\\
	\braket{\Psi_{\vec{n}_A^{\prime}}|C_{\vec{n}_A^{\prime}}\hat{H}_{AB}^{(v2)}C^{\dagger}_{\vec{n}_A}|\Psi_{\vec{n}_A}}=&\sum_{\mu\mu'\in\boldsymbol{\mathrm M}_A}\sum_{\nu\nu'\in\boldsymbol{\mathrm M}_B}\Big[\big(v_{\mu\nu\mu'\nu'}-v_{\mu\nu\nu'\mu'}-v_{\nu\mu\mu'\nu'}+v_{\nu\mu\nu'\mu'}\big)\braket{\vec{n}_A^{\prime}|c^{\dagger}_\mu c_{\mu'}|\vec{n}_A}\braket{\Psi_{\vec{n}_A^{\prime}}|c^{\dagger}_\nu c_{\nu'}|\Psi_{\vec{n}_A}}\nonumber\\
	&+v_{\mu\mu'\nu\nu'}\braket{\vec{n}_A^{\prime}|c^{\dagger}_\mu c^{\dagger}_{\mu'}|\vec{n}_A}\braket{\Psi_{\vec{n}_A^{\prime}}|c_{\nu'}c_\nu|\Psi_{\vec{n}_A}}+v_{\nu'\nu\mu'\mu}\braket{\vec{n}_A^{\prime}|c_\mu c_{\mu'}|\vec{n}_A}\braket{\Psi_{\vec{n}_A^{\prime}}|c^{\dagger}_{\nu'}c^{\dagger}_\nu|\Psi_{\vec{n}_A}}\Big],\\
	\braket{\Psi_{\vec{n}_A^{\prime}}|C_{\vec{n}_A^{\prime}}\hat{H}_{AB}^{(v3)}C^{\dagger}_{\vec{n}_A}|\Psi_{\vec{n}_A}}=&\sum_{\mu\in\boldsymbol{\mathrm M}_A}\sum_{\mu'\nu\nu'\in\boldsymbol{\mathrm M}_B}(-1)^{N_A}\Big[\big(v_{\mu\nu\mu'\nu'}-v_{\nu\mu\mu'\nu'}\big)\braket{\vec{n}_A^{\prime}|c^{\dagger}_\mu|\vec{n}_A}\braket{\Psi_{\vec{n}_A^{\prime}}|c^{\dagger}_\nu c_{\nu'}c_{\mu'}|\Psi_{\vec{n}_A}}\nonumber\\
	&+\big(v_{\nu'\nu\mu'\mu}-v_{\nu'\nu\mu\mu'}\big)\braket{\vec{n}_A^{\prime}|c_{\mu}|\vec{n}_A}\braket{\Psi_{\vec{n}_A^{\prime}}|c^{\dagger}_{\nu'} c^{\dagger}_\nu c_{\mu'}|\Psi_{\vec{n}_A}}\Big].
\end{align}
\end{widetext}

Every factor of the form $\braket{\vec{n}^{\prime}_A|\hat{O}_A|\vec{n}_{A}}$, where $\hat{O}_A$ is some operator acting on states in subspace $A$, can be expressed as a Kronecker delta function eliminating all but one term in the sum over $\vec{n}_A^{\prime}$ in Eq.~(\ref{eq:expectationH}).  These delta functions will be found classically.  The other type of factors of the form $\braket{\Psi_{\vec{n}_A^{\prime}}|\hat{O}_B|\Psi_{\vec{n}_A}}$, where $\hat{O}_B$ is some operator acting on states in subspace $B$, is computed using our modified VQE algorithm.

First, let us require that the mapping from $\mathcal F_B$ to the Hilbert space for the quantum register takes $\ket{\Psi_{\vec{n}_A}}$ to $\ket{\Phi_{\vec{n}_A}}$ for all configurations $\vec{n}_A$.  Moreover, if we define $\hat{\mathcal O}$ such that $\hat{O}_B\mapsto\hat{\mathcal O}$, we find that the factors $\braket{\Psi_{\vec{n}_A^{\prime}}|\hat{O}_B|\Psi_{\vec{n}_A}}$ map to $\braket{\Phi_{\vec{n}^{\prime}_A}|\hat{\mathcal O}|\Phi_{\vec{n}_A}}$.  In the case, $\vec{n}_A^{\prime}=\vec{n}_A$, the latter become expectation values that can be determined using the standard VQE algorithm by preparing the state $\ket{\Phi_{\vec{n}_A}}$ on the quantum register and performing appropriate measurements.  For $\vec{n}_A^{\prime}\ne\vec{n}_A$, this approach is not possible, so we instead introduce an ancilla qubit and prepare the state
\begin{equation}
    \ket{\Phi_{\vec{n}_A\vec{n}^{\prime}_A}} = \frac{1}{\sqrt{2}}\Big(\ket{0;\Phi_{\vec{n}_A}} + \ket{1;\Phi_{\vec{n}^{\prime}_A}}\Big)
    \label{psinn}
\end{equation}
on the quantum register.  This allows us to determine
\begin{equation}
	\braket{\Phi_{\vec{n}^{\prime}_A}|\hat{\mathcal O}|\Phi_{\vec{n}_A}}=\braket{\Phi_{\vec{n}_A\vec{n}^{\prime}_A}|(X-i Y)\otimes\hat{\mathcal O}|\Phi_{\vec{n}_A\vec{n}^{\prime}_A}},
	\label{psinn2}
\end{equation}
where $X$ and $Y$ are Pauli operators, by computing the latter expectation value using the same technique as in the standard VQE algorithm for the non-ancilla qubits.

\subsection{VQE on blocks}

In standard VQE, the ansatz is a parameterized unitary operator $\hat U(\vec{\theta})$, which when acting on the initial state $\ket{\vec 0} = \ket{00\ldots0}$ of the quantum register produce $\ket{\Phi}$ representing the many-particle state $\ket{\Psi}\in\mathcal F$.  Herein, we instead want to represent $\ket{\Psi_{\vec{n}_A}}\in\mathcal F_B$.  As the states on subsystem $B$ are dependent on the particle configuration of subsystem $A$, we are interested in the $\vec{n}_A$-dependent ansatz states $\hat U(\vec{\theta}_{\vec{n}_{A}})\ket{\vec 0}$ to describe the quantum states
\begin{equation}
    \ket{\Phi_{\vec{n}_{A}}} \approx \hat U(\vec{\theta}_{\vec{n}_{A}})\ket{\vec 0}
\end{equation}
on the quantum register corresponding to $\ket{\Psi_{\vec{n}_A}}$.  The energy expectation values can be calculated as described in sec~\ref{GMSS}.  These values are then used to optimize, not only each $\vec{\theta}_{\vec{n}_{A}}$ but also each $\alpha_{\vec{n}_{A}}$ subject to the normalization constraint $\sum_{\vec{n}_{A}} |\alpha_{\vec{n}_{A}}|^2 = 1$.  As noted above, to calculate
\begin{equation}
    \braket{\Phi_{\vec{n}^{\prime}_{A}}|\hat{\mathcal O}|\Phi_{\vec{n}_{A}}}\approx\braket{\vec 0|\hat U^{\dagger}(\vec{\theta}_{\vec{n}_{A}'})\hat{\mathcal O}\hat U(\vec{\theta}_{\vec{n}_{A}})|\vec 0}
\end{equation}
we use an ancilla qubit to form the state $\ket{\Phi_{\vec{n}_{A}\vec{n}_{A}'}}$ in Eq.~\eqref{psinn}.

\begin{figure}[t]
\begin{center}
\includegraphics[width=\columnwidth]{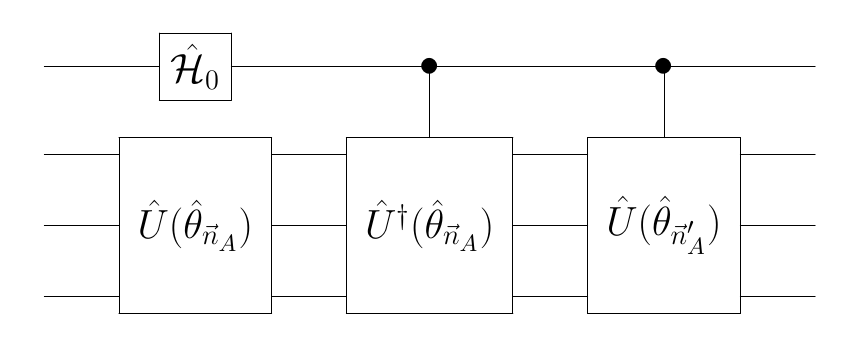}
\end{center}
\vspace{-2mm}

\caption{Unitary which creates the $\ket{+;\Phi_{\vec{n}_{A}\vec{n}_{A}'}}$ state used for calculating off diagonal brackets.  In this example, each block is described by three qubits, however, the unitary can easily be generalized to any number of qubits.}
\label{FN}
\vspace{3mm}
\end{figure}

What we have not yet addressed is how to create this state using quantum gates.  Specifically, we need to implement a controlled unitary operator $\hat C_{\hat U(\vec{\theta}_{\vec{n}_{A}})}$ that applies $\hat U(\vec{\theta}_{\vec{n}_{A}})$, if and only if the ancilla qubit is in the state $\ket{1}$.  We can then obtain the desired state by using
\begin{equation}
    \ket{+;\Phi_{\vec{n}_{A}\vec{n}_{A}'}} \approx \hat C_{\hat U(\vec{\theta}_{\vec{n'}_{A}})}\hat C_{\hat U^{\dagger}(\vec{\theta}_{\vec{n}_{A}})}\hat [\mathcal{H}_0\otimes U(\vec{\theta}_{\vec{n}_{A}})]\ket{0;\vec 0}
    \label{psicc_u}
\end{equation}
where $\ket{+}=\big(\ket{0}+\ket{1}\big)/\sqrt{2}$ and $\hat{\mathcal{H}_0}$ is the Hadamard gate acting on the ancilla qubit, see Fig~\ref{FN}.  The $\hat C_{\hat U(\vec{\theta})}$ gates can be constructed using $C_{\text{NOT}}$ gates.  However, $C_{\text{NOT}}$ gates are a major source of error.  Since often $|\vec{\theta}_{\vec{n}_{A}}|\ll\pi$, we use scaled two-qubit cross-resonance gates which are significantly more efficient at small angles~\cite{stenger2021}.

\section{Hubbard Model}

To demonstrate our approach, we focus below on the $L$-site Hubbard model of a system with identical spin-$1/2$ particles described by the Hamiltonian
\begin{equation}
   \hat{H} = \epsilon \sum_{\sigma} \hat{N}_{\sigma} + t \sum_{\sigma}\hat{T}_\sigma  +U \sum_i \hat{n}_{i\uparrow}\hat{n}_{i\downarrow},
	\label{hubbard}
\end{equation}
where 
\begin{subequations}
\begin{align}
	\hat N_\sigma &= \sum_i\hat n_{i\sigma}, \\
	\hat T_\sigma &= \sum_i\left(c^{\dagger}_{i\sigma}c_{i+1\sigma}+\text{h.c.}\right),
\end{align} 
\end{subequations}%
and the fermionic operators, including the number operators $\hat{n}_{i\sigma} = c^{\dagger}_{i\sigma}c_{i\sigma}$, are identified by single-particle indices split into site indices $i\in\{1,2,\ldots,L\}$ and spin indices $\sigma\in\{\uparrow,\downarrow\}$, $\epsilon$ is the electrochemical potential, and $t$ and $U$ are the hopping and onsite Coulomb interaction coefficients, respectively.  The sites are arranged in a one-dimensional ring with the boundary condition $c_{L\sigma} = c_{0\sigma}$.

\subsection{Splitting the spin sectors}

Placing spin-up and spin-down particles in subsystems $A$ and $B$, respectively, allows us to identify the particle configurations as $\vec{n}_{\sigma}=\big(n_{1\sigma},n_{2\sigma},\ldots,n_{L\sigma}\big)$ and write the many-particle state as
\begin{equation}
    \ket{\Psi} = \sum_{\vec{n}_\uparrow\vec{n}_\downarrow}\alpha_{\vec{n}_{\uparrow}}\beta_{\vec{n}_{\uparrow}\vec{n}_{\downarrow}}\ket{\vec{n}},
\end{equation}
where the normalization yields $\sum_{\vec{n}_{\uparrow}}|\alpha_{\vec{n}_{\uparrow}}|^2 = 1$ and $\sum_{\vec{n}_{\downarrow}}|\beta_{\vec{n}_{\uparrow}\vec{n}_{\downarrow}}|^2 = 1$, the latter for all $\vec{n}_{\uparrow}$.  Without any restrictions on the vector spaces for the two subsystems, the number of spin-up states and the number of spin-down states are both $2^L$.

In order to perform VQE we need to calculate the energy expectation value which is used as the cost function.  We will calculate the spin-up sector of the expectation value classically.  This will leave us with a set of $\vec{n}_{\uparrow}$-dependent spin-down states
\begin{equation}
    \ket{\Psi_{\vec{n}_{\uparrow}}} = \sum_{\vec{n}_{\downarrow}}\beta_{\vec{n}_{\uparrow}\vec{n}_\downarrow}\ket{\vec{n}_{\downarrow}}
\end{equation}
in the Fock space $\mathcal F_\downarrow$.  We then use the quantum device to determine the coefficients $\beta_{\vec{n}_{\uparrow}\vec{n}_\downarrow}$.  The expectation value of the Hamiltonian can be expressed as
\begin{align}
\bra{\Psi}\hat{H}\ket{\Psi} 
 =&~\epsilon\sum_{\vec{n}_{\uparrow}}|\alpha_{\vec{n}_{\uparrow}}|^2\Big(N_{\vec{n}_{\uparrow}} + \braket{\Psi_{\vec{n}_{\uparrow}}|\hat{N}_{\downarrow}|\Psi_{\vec{n}_{\uparrow}}}\Big)\nonumber\\
 &+t\sum_{\vec{n}_{\uparrow}\vec{n}_{\uparrow}^{\prime}}\alpha_{\vec{n}_{\uparrow}^{\prime}}^*\alpha_{\vec{n}_{\uparrow}}  T_{\vec{n}_{\uparrow}^{\prime}\vec{n}_{\uparrow}}\braket{\Psi_{\vec{n}_{\uparrow}^{\prime}}|\Psi_{\vec{n}_{\uparrow}}} \nonumber\\
 &+t\sum_{\vec{n}_{\uparrow}}|\alpha_{\vec{n}_{\uparrow}}|^2 \braket{\Psi_{\vec{n}_{\uparrow}}|\hat{T}_{\downarrow}|\Psi_{\vec{n}_{\uparrow}}} \nonumber\\
& +U\sum_{\vec{n}_{\uparrow}}|\alpha_{\vec{n}_{\uparrow}}|^2  \sum_i  n_{i\uparrow}\braket{\Psi_{\vec{n}_{\uparrow}}|\hat{n}_{i\downarrow}|\Psi_{\vec{n}_{\uparrow}}},
\label{Hblocks}
\end{align}
where $N_{\vec{n}_\uparrow} = \bra{\vec{n}_{\uparrow}}\hat{N}_{\uparrow}\ket{\vec{n}_{\uparrow}}$ and $T_{\vec{n}_{\uparrow}^{\prime}\vec{n}_{\uparrow}} = \bra{\vec{n}_{\uparrow}^{\prime}}\hat{T}_{\uparrow}\ket{\vec{n}_{\uparrow}}$.  The spin-up factors $n_{i\uparrow}$, $N_{\vec{n}_\uparrow}$, and $T_{\vec{n}_{\uparrow}^{\prime}\vec{n}_{\uparrow}}$ will be calculated on the classical computer and stored for use in our modified VQE.  The spin-down factors $\braket{\Psi_{\vec{n}_{\uparrow}'}|\Psi_{\vec{n}_{\uparrow}}}$, $\braket{\Psi_{\vec{n}_{\uparrow}}|\hat{T}_{\downarrow}|\Psi_{\vec{n}_{\uparrow}}}$, and $\braket{\Psi_{\vec{n}_{\uparrow}}|\hat{n}_{i\downarrow}|\Psi_{\vec{n}_{\uparrow}}}$ will be calculated using the quantum computer.

The expectation values $\braket{\Psi_{\vec{n}_{\uparrow}}|\hat{T}_{\downarrow}|\Psi_{\vec{n}_{\uparrow}}}$, and $\braket{\Psi_{\vec{n}_{\uparrow}}|\hat{n}_{i\downarrow}|\Psi_{\vec{n}_{\uparrow}}}$ are straightforward to calculate using the quantum device.  First, map fermonic states $\ket{\Psi_{\vec{n}_{\uparrow}}}$ onto states of the quantum computer $\ket{\Phi_{\vec{n}_{\uparrow}}}$ and decompose $\hat{T}_{\downarrow}$ and $\hat{n}_{i\downarrow}$ into Pauli operators.  Then measure the expectation value of the Pauli matrices using standard techniques.  The bracket $\braket{\Psi_{\vec{n}_{\uparrow}'}|\Psi_{\vec{n}_{\uparrow}}}$ is harder to calculate since it does not have the form of an expectation value.  For this bracket we add an ancilla qubit as in Eqns~\eqref{psinn} and~\eqref{psinn2}.

By calculating the spin up sector on the classical computer, we have reduced the number of qubits required for the quantum computation from $2L$ to $L + 1$.  An additional advantage of splitting between the spin sectors is that it exploits the sparsity of the Hamiltonian.  Everything is diagonal in the spin-up sector except for the elements $T_{\vec{n}_{\uparrow}^{\prime}\vec{n}_{\uparrow}}$.  Furthermore, $T_{\vec{n}_{\uparrow}^{\prime}\vec{n}_{\uparrow}}$ is sparse with many blocks having all elements equal to zero.  Because we calculate the $T_{\vec{n}_{\uparrow}^{\prime}\vec{n}_{\uparrow}}$ terms before running the spin-down sector on the quantum processor, we can 
ensure that only blocks with non-zero elements are run on the quantum device. 

The price of calculating the spin-up sector classically is that we must sum over $2^L$ spin-up configurations.  This number can be reduced, however, by restricting the total number spin-up fermions.

\subsection{Fixing the number of spin-up fermions}
\label{FSU}

The Hubbard model, Eq.~\eqref{hubbard}, conserves the total number of spin up and spin down fermions.  Therefore, we can reduce the size of the classical computation by fixing the number of spin-up electrons  $\hat{N}_{\uparrow} \rightarrow N_{\uparrow}$.  In general there are
\begin{equation}
   \Lambda = {L \choose N_{\uparrow}}
\end{equation}
spin up states.  

This procedure still requires $Q=L + 1$ qubits to solve the down spins.  However, the complexity of the classical computation is greatly reduced.  There is no disadvantage to fixing the number of spin-up fermions.

\subsection{Fixing the spin-down parity}

A single qubit can be removed by fixing the total parity of the spin-down sector~\cite{bravyi2017}.  This can be done by changing the qubit encoding of our fermion operators.  A natural choice for for encoding is the Jordan--Wigner transformation~\cite{jordan1928} for which 
\begin{equation}
    c_{i\downarrow}  \mapsto \frac{1}{2}(X_i  + i Y_i )\prod_{j=0}^{L-1}Z_j  
\end{equation}
where $X_i$, $Y_i$, and $Z_i$ are the three Pauli matrices acting on qubit $i$.  This is convenient because of the simplicity of interpretation, a fermion operator at site $i$ flips the state of qubit $i$.  However, it uses more qubits than is necessary as fermion parity is always conserved.

We can remove a qubit by changing the encoding to, for example, the parity encoding~\cite{Seeley2012,Bravyi2002}.  Here the fermion operators are given by
\begin{equation}
        c_{i\downarrow}  \mapsto \frac{1}{2}(Z_{i-1}X_i  + i Y_i)\prod_{j=i+1}^{L}X_j 
\end{equation}
Note that in this encoding the total parity maps to qubit zero.  Therefore, we can remove this qubit from the encoding and set the value of the parity by hand.  The downside is that the Pauli decomposition  of $\hat{T}_{\downarrow}$ and $\hat{n}_{i\downarrow}$ contains a greater number of Pauli terms.  For the most part, the Pauli-strings only grow by one qubit, however, the boundary term in $\hat{T}_{\downarrow}$ (i.e. $c^{\dagger}_{L-1\downarrow}c_{0\downarrow} +\text{h.c.}$) is mapped onto Pauli-strings of length $L$.  

This situation can be improved by selecting yet another encoding, namely the Bravyi-Kitaev encoding~\cite{Bravyi2002,Seeley2012}.  The Bravyi-Kitaev transformation balances information about locality with information about parity so that the length of the Pauli-strings for local fermion operators is $\mathcal{O}(\log L)$ while also having the property that the total parity is mapped to the first qubit.  The exact form of the transformation is complicated and can be found in the literature~\cite{Bravyi2002,Seeley2012,bravyi2017} so we will not write it here.  For the one-dimensional Hubbard model, the Pauli-strings are still larger on average than in the Jordan--Wigner transformation.  One has to make a hardware-based decision for whether it is better to have fewer qubits or shorter Pauli-strings.    

\subsection{Fixing the number of spin down fermions}
\label{FSUASD}

Since the Hubbard model conserves the number of spin-down electrons as well as the number of spin-up electrons, we can also fix the total number of spin down fermions to $\hat{N}_{\downarrow} \rightarrow N_{\downarrow}$.  By fixing the total number of spin-down electrons, we reduce the Hilbert space of $\hat{T}_{\downarrow}$ and $\hat{n}_{i\downarrow}$ which means we can write $\ket{\Phi_{\vec{n}_{\uparrow}}}$ using fewer qubits from $Q=L + 1$ to
\begin{equation}
    Q = \bigg\lceil \log_2 { L \choose N_{\downarrow}}\bigg\rceil + 1 
    \label{Qnum}
\end{equation}
This can be a major gain, however, it does increase the amount of work required for the classical computer.  In fact, the number of Pauli matrices required to describe $\hat{T}_{\downarrow}$ is no longer guaranteed to be linear with system size.  

To find the Pauli decomposition of $\hat{T}_{\downarrow}$ and $\hat{n}_{i\downarrow}$ we have to first create the matrix representation of both in the basis were both spins are fixed.  This step is done classically and can be done efficiently by calling the states using the binary representation of integers then we can find matrix elements using binary operations, see \cite{Lin1993} for details.  Once the matrix has been specified, the Pauli coefficients can be found in $\mathcal{O}\left(L^2 \log_2 L\right)$ operations, or less~\cite{gunlycke2020}.  

\begin{figure}[t]
\begin{center}
\includegraphics[width=\columnwidth]{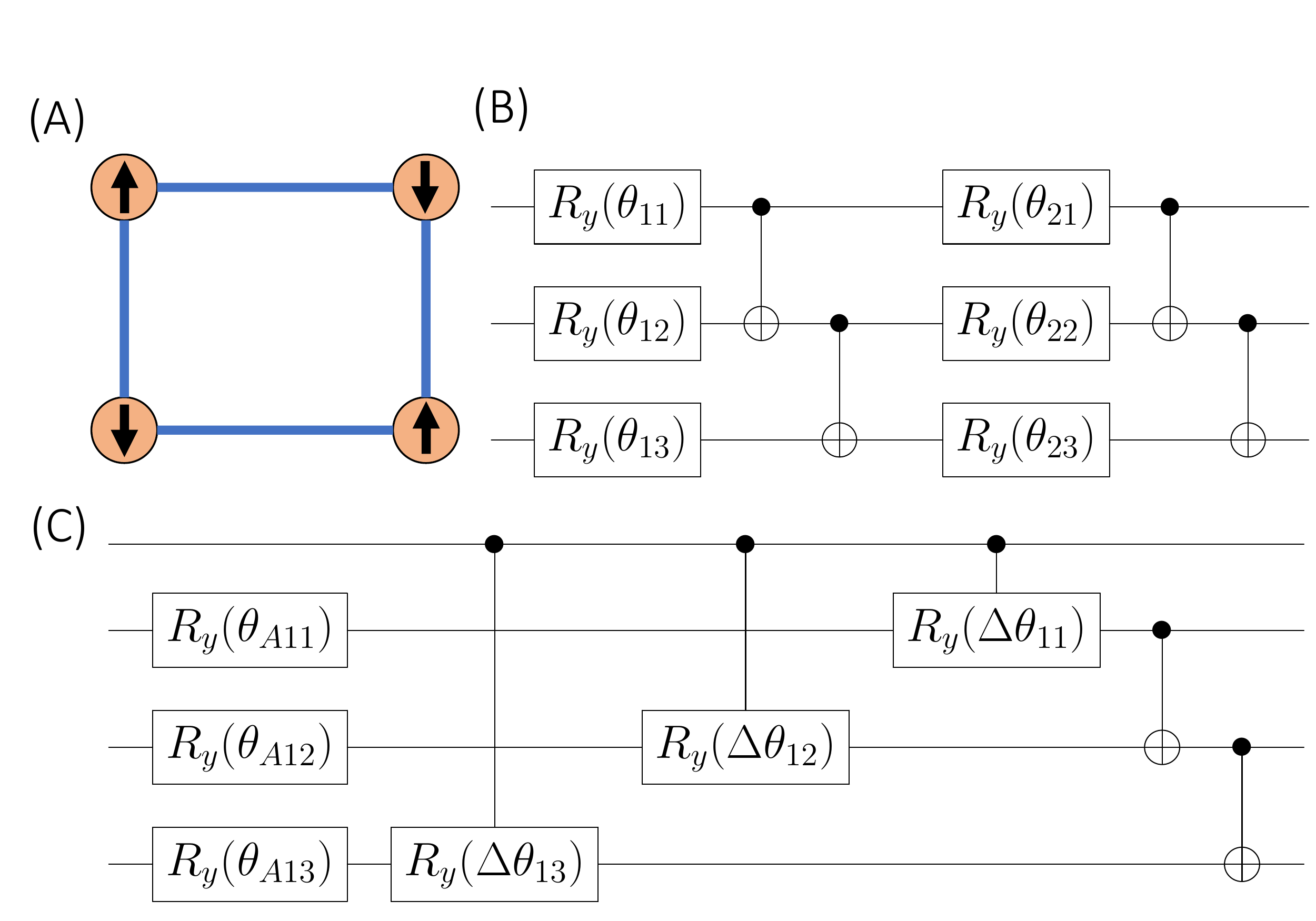}
\end{center}
\vspace{-2mm}

\caption{Simulation details.  (A) depiction of the four site Hubbard Model at half filling.  The orange circles represent sites and the blue lines show connectivity.  The arrows represent spin-up and spin-down electrons.  The placement of the electrons in the figure does not necessarily reflect the actual location of electrons.   (B) two iterations of the ansatz used for our simulations. The single qubit rotation during iteration $n$ on qubit $q$ is parameterized by $\theta_{nq}$.  In the simulation we use two iterations of the ansatz. (C) circuit to apply a single iteration of the ansatz to off diagonal blocks.  The set of rotations on block $A$ are $\{ \theta_{A nq} \}$ and those on block $B$ are $\{ \theta_{B nq} \}$.  The difference between two phases from different blocks is labeled as $\Delta\theta_{nq} = \theta_{Bnq} - \theta_{Anq}$.  This circuit applies the block $A$ rotations if the ancilla qubit is off and the block $B$ rotations if the ancilla qubit is on. The controlled-$R_y$ gates are generated using parameterized $R_{ZX}$ gates~\cite{stenger2021}.}
\label{F0}
\vspace{3mm}
\end{figure}

\section{Simulation of the four site Hubbard model at half filling fixing the number of both spin sectors}

In this section, we present results for a simulation of the four site ($L=4$) Hubbard model [Eq.~\eqref{hubbard}] at half filling ($N_{\uparrow} = N_{\downarrow} = 2$).  See Fig.~\ref{F0}(A).  The number of spin-up states is $[L ~ \text{choose} ~ N_{\uparrow}] = 6$.  The number of qubits we need is $Q = 4$ (see Eq.~\eqref{Qnum}).

We use the hardware-efficient ansatz~\cite{Kandala2017,Fischer2019} where single qubit gates are sandwiched between CNOT gates.  See Fig.~\ref{F0}(B).  We use two iterations of the ansatz for our simulation.

\begin{figure}[t]
\begin{center}
\vspace{4mm}
\includegraphics[width=\columnwidth]{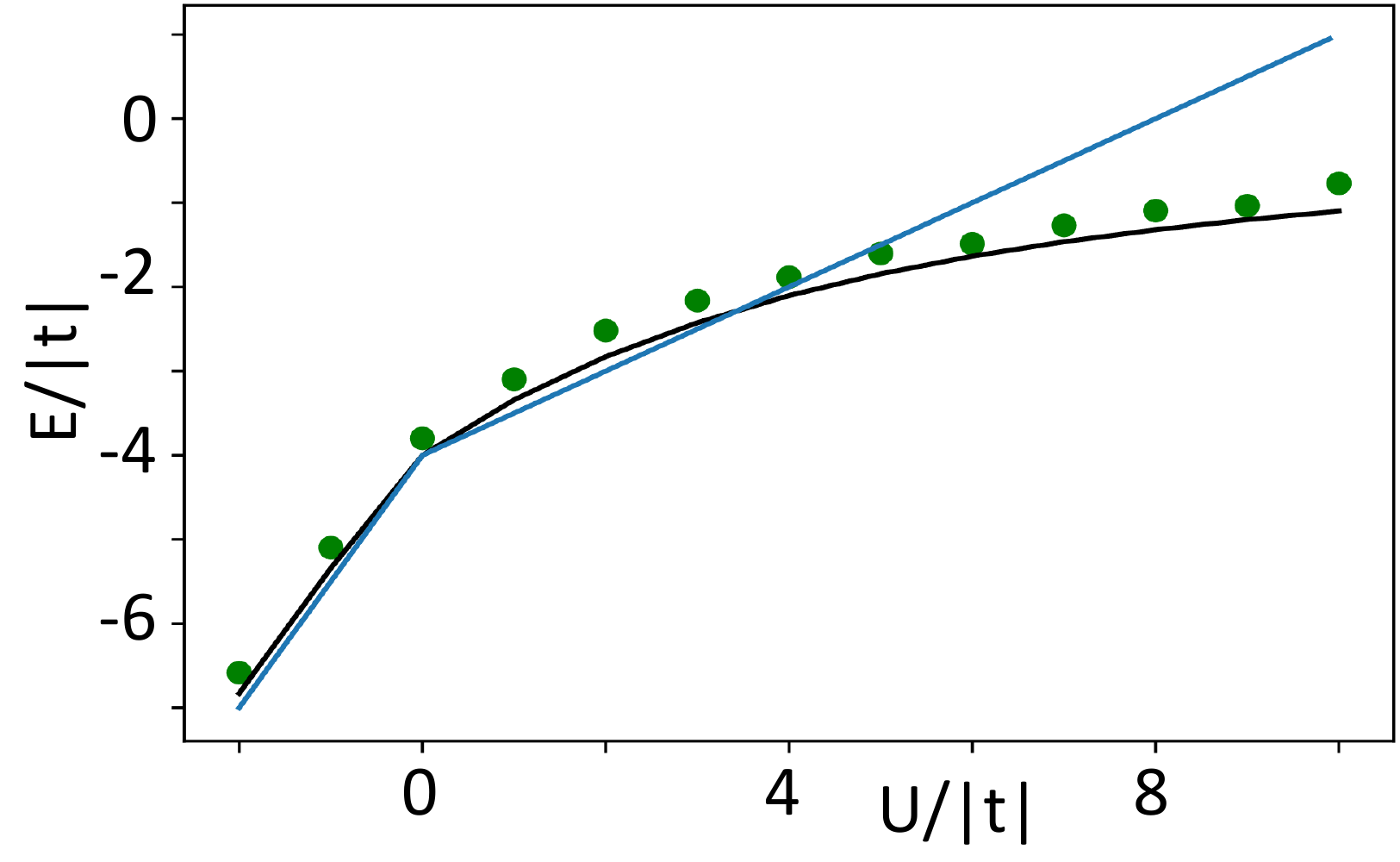}
\end{center}
\vspace{-2mm}

\caption{Results from the simulation of our block VQE method on the four site Hubbard model at half filling, fixing the number of particles in both spin sectors.  The results of block VQE are plotted as green circles.  The exact ground state is plotted as the solid black curve.  The mean field result is plotted as the solid blue curve.}
\label{F1}
\vspace{3mm}
\end{figure}

In order to calculate the off diagonal blocks, we add an ancilla qubit.  For the particular ansatz we use, the general form of the off diagonal unitary in Eq.~\ref{psicc_u} can be simplified, see Fig~\ref{F0}(C).

The results of the simulations are plotted as the green circles in Fig.~\ref{F1}.  The results are compared against the exact ground state (solid black curve) and the mean field solution (solid blue curve).  For large values of the interaction strength $U$, our VQE is much better than the mean field calculation.



\section{Conclusion}

We have described a method which splits the work of solving many-particle Hamiltonians between classical and quantum computers.  In our main example, this is accomplished by splitting the spin-sectors and solving one of them classically before solving the other on the quantum device.  Splitting the spins in this way immediately reduces the number of qubits needed for the quantum calculation by half.  A further reduction of qubits can be achieved by evoking parity, and even more qubits can be removed by invoking spin number conservation.  Each time qubits are removed from the quantum calculation, the classical calculation becomes more involved.  Thus, we propose a ladder of strategies going from most classically intensive and least quantum intensive to least classically intensive and most quantum intensive.  As quantum processors improve we will be able to climb up this ladder.    

\appendix*
\section{System specification}

The coefficients describing the system can be computed using the integral forms:
\begin{equation}
	t_{\mu\mu'}=\int \psi_\mu^*(x)\hat T(x)\psi_{\mu'}(x)\mathrm{d}x
\end{equation}
and
\begin{equation}
	v_{\mu\nu\mu'\nu'}=\iint \psi_\mu^*(x)\psi_\nu^*(x')\hat V(x,x')\psi_{\mu'}(x)\psi_{\nu'}(x')\mathrm{d}x\mathrm{d}x',
\end{equation}
where $\psi_\mu(x)$ are one-particle basis functions on some coordinate space and $\hat T$ and $\hat V$ are first-quantized one- and two-particle operators, respectively.

For an isolated molecular system within the Born--Oppenheimer approximation, the integral coefficients for the electronic system can be expressed as
\begin{widetext}
\begin{align}
	t_{\mu\mu'}&=\sum_\sigma\int \psi_\mu^*(\vec r,\sigma)\Bigg[-\frac{\hbar^2}{2m}\nabla^2-\frac{e^2}{4\pi\varepsilon_0}\sum_n\frac{Z_n}{\vec r-\vec R_n}\Bigg]\psi_{\mu'}(\vec r,\sigma)\,\mathrm{d}\vec r,\\
	v_{\mu\nu\mu'\nu'}&=\sum_{\sigma\sigma'}\iint \psi_\mu^*(\vec r,\sigma)\psi_\nu^*(\vec r',\sigma')\frac{e^2}{4\pi\varepsilon_0|\vec r-\vec r'|}\psi_{\mu'}(\vec r,\sigma)\psi_{\nu'}(\vec r',\sigma')\,\mathrm{d}\vec r\,\mathrm{d}\vec r'	
\end{align}
\end{widetext}
where $\vec r$ and $\sigma$ are spatial and spin coordinates, respectively, $\hbar$, $m$, $e$, and $\epsilon_0$ are the reduced Planck constant, the electron mass, the electron charge, and the vacuum permittivity, respectively, and $Z_n$ and $\vec R_n$ are the charges and locations of the nuclei $n$.

\begin{acknowledgments}

This work has been supported by the Office of Naval Research (ONR) through the U.S. Naval Research Laboratory (NRL).  J.P.T.S. thanks the National Research Council Research Associateship Programs.  We acknowledge quantum computing resources from IBM through a collaboration with the Air Force Research Laboratory (AFRL).

\end{acknowledgments}

\bibliography{references}

\begin{thebibliography}{31}%
\makeatletter
\providecommand \@ifxundefined [1]{%
 \@ifx{#1\undefined}
}%
\providecommand \@ifnum [1]{%
 \ifnum #1\expandafter \@firstoftwo
 \else \expandafter \@secondoftwo
 \fi
}%
\providecommand \@ifx [1]{%
 \ifx #1\expandafter \@firstoftwo
 \else \expandafter \@secondoftwo
 \fi
}%
\providecommand \natexlab [1]{#1}%
\providecommand \enquote  [1]{``#1''}%
\providecommand \bibnamefont  [1]{#1}%
\providecommand \bibfnamefont [1]{#1}%
\providecommand \citenamefont [1]{#1}%
\providecommand \href@noop [0]{\@secondoftwo}%
\providecommand \href [0]{\begingroup \@sanitize@url \@href}%
\providecommand \@href[1]{\@@startlink{#1}\@@href}%
\providecommand \@@href[1]{\endgroup#1\@@endlink}%
\providecommand \@sanitize@url [0]{\catcode `\\12\catcode `\$12\catcode
  `\&12\catcode `\#12\catcode `\^12\catcode `\_12\catcode `\%12\relax}%
\providecommand \@@startlink[1]{}%
\providecommand \@@endlink[0]{}%
\providecommand \url  [0]{\begingroup\@sanitize@url \@url }%
\providecommand \@url [1]{\endgroup\@href {#1}{\urlprefix }}%
\providecommand \urlprefix  [0]{URL }%
\providecommand \Eprint [0]{\href }%
\providecommand \doibase [0]{https://doi.org/}%
\providecommand \selectlanguage [0]{\@gobble}%
\providecommand \bibinfo  [0]{\@secondoftwo}%
\providecommand \bibfield  [0]{\@secondoftwo}%
\providecommand \translation [1]{[#1]}%
\providecommand \BibitemOpen [0]{}%
\providecommand \bibitemStop [0]{}%
\providecommand \bibitemNoStop [0]{.\EOS\space}%
\providecommand \EOS [0]{\spacefactor3000\relax}%
\providecommand \BibitemShut  [1]{\csname bibitem#1\endcsname}%
\let\auto@bib@innerbib\@empty
\bibitem [{\citenamefont {Feynman}(1982)}]{Feynman1982}%
  \BibitemOpen
  \bibfield  {author} {\bibinfo {author} {\bibfnamefont {R.~P.}\ \bibnamefont
  {Feynman}},\ }\href {https://doi.org/10.1007/BF02650179} {\bibfield
  {journal} {\bibinfo  {journal} {Int. J. Theor Phys.}\ }\textbf {\bibinfo
  {volume} {21}},\ \bibinfo {pages} {467} (\bibinfo {year} {1982})}\BibitemShut
  {NoStop}%
\bibitem [{\citenamefont {Lloyd}(1996)}]{Lloyd1996}%
  \BibitemOpen
  \bibfield  {author} {\bibinfo {author} {\bibfnamefont {S.}~\bibnamefont
  {Lloyd}},\ }\href {https://doi.org/10.1126/science.273.5278.1073} {\bibfield
  {journal} {\bibinfo  {journal} {Science (New York, N.Y.)}\ }\textbf {\bibinfo
  {volume} {273}},\ \bibinfo {pages} {1073} (\bibinfo {year}
  {1996})}\BibitemShut {NoStop}%
\bibitem [{\citenamefont {Boixo}\ \emph {et~al.}(2018)\citenamefont {Boixo},
  \citenamefont {Isakov}, \citenamefont {Smelyanskiy}, \citenamefont {Babbush},
  \citenamefont {Ding}, \citenamefont {Jiang}, \citenamefont {Bremner},
  \citenamefont {Martinis},\ and\ \citenamefont {Neven}}]{Boixo2018}%
  \BibitemOpen
  \bibfield  {author} {\bibinfo {author} {\bibfnamefont {S.}~\bibnamefont
  {Boixo}}, \bibinfo {author} {\bibfnamefont {S.~V.}\ \bibnamefont {Isakov}},
  \bibinfo {author} {\bibfnamefont {V.~N.}\ \bibnamefont {Smelyanskiy}},
  \bibinfo {author} {\bibfnamefont {R.}~\bibnamefont {Babbush}}, \bibinfo
  {author} {\bibfnamefont {N.}~\bibnamefont {Ding}}, \bibinfo {author}
  {\bibfnamefont {Z.}~\bibnamefont {Jiang}}, \bibinfo {author} {\bibfnamefont
  {M.~J.}\ \bibnamefont {Bremner}}, \bibinfo {author} {\bibfnamefont {J.~M.}\
  \bibnamefont {Martinis}},\ and\ \bibinfo {author} {\bibfnamefont
  {H.}~\bibnamefont {Neven}},\ }\href
  {https://doi.org/10.1038/s41567-018-0124-x} {\bibfield  {journal} {\bibinfo
  {journal} {Nature Physics}\ }\textbf {\bibinfo {volume} {14}},\ \bibinfo
  {pages} {595–600} (\bibinfo {year} {2018})}\BibitemShut {NoStop}%
\bibitem [{\citenamefont {Shor}(1997)}]{Shor1997}%
  \BibitemOpen
  \bibfield  {author} {\bibinfo {author} {\bibfnamefont {P.~W.}\ \bibnamefont
  {Shor}},\ }\href {https://doi.org/10.1137/s0097539795293172} {\bibfield
  {journal} {\bibinfo  {journal} {SIAM Journal on Computing}\ }\textbf
  {\bibinfo {volume} {26}},\ \bibinfo {pages} {1484–1509} (\bibinfo {year}
  {1997})}\BibitemShut {NoStop}%
\bibitem [{\citenamefont {Cirac}\ and\ \citenamefont
  {Zoller}(2012)}]{Cirac2012}%
  \BibitemOpen
  \bibfield  {author} {\bibinfo {author} {\bibfnamefont {J.~I.}\ \bibnamefont
  {Cirac}}\ and\ \bibinfo {author} {\bibfnamefont {P.}~\bibnamefont {Zoller}},\
  }\href {https://doi.org/10.1038/nphys2275} {\bibfield  {journal} {\bibinfo
  {journal} {nature physics}\ }\textbf {\bibinfo {volume} {8}},\ \bibinfo
  {pages} {264–266} (\bibinfo {year} {2012})}\BibitemShut {NoStop}%
\bibitem [{\citenamefont {Georgescu}\ \emph {et~al.}(2014)\citenamefont
  {Georgescu}, \citenamefont {Ashhab},\ and\ \citenamefont
  {Nori}}]{Georgescu2014}%
  \BibitemOpen
  \bibfield  {author} {\bibinfo {author} {\bibfnamefont {I.}~\bibnamefont
  {Georgescu}}, \bibinfo {author} {\bibfnamefont {S.}~\bibnamefont {Ashhab}},\
  and\ \bibinfo {author} {\bibfnamefont {F.}~\bibnamefont {Nori}},\ }\href
  {https://doi.org/10.1103/revmodphys.86.153} {\bibfield  {journal} {\bibinfo
  {journal} {Reviews of Modern Physics}\ }\textbf {\bibinfo {volume} {86}},\
  \bibinfo {pages} {153–185} (\bibinfo {year} {2014})}\BibitemShut {NoStop}%
\bibitem [{\citenamefont {Tarruell}\ and\ \citenamefont
  {Sanchez-Palencia}(2019)}]{tarruell2019}%
  \BibitemOpen
  \bibfield  {author} {\bibinfo {author} {\bibfnamefont {L.}~\bibnamefont
  {Tarruell}}\ and\ \bibinfo {author} {\bibfnamefont {L.}~\bibnamefont
  {Sanchez-Palencia}},\ }\href@noop {} {\  (\bibinfo {year} {2019})},\ \Eprint
  {https://arxiv.org/abs/1809.00571} {arXiv:1809.00571 [cond-mat.quant-gas]}
  \BibitemShut {NoStop}%
\bibitem [{\citenamefont {Salfi}\ \emph {et~al.}(2016)\citenamefont {Salfi},
  \citenamefont {Mol}, \citenamefont {Rahman}, \citenamefont {Klimeck},
  \citenamefont {Simmons}, \citenamefont {Hollenberg},\ and\ \citenamefont
  {Rogge}}]{Salfi2016}%
  \BibitemOpen
  \bibfield  {author} {\bibinfo {author} {\bibfnamefont {J.}~\bibnamefont
  {Salfi}}, \bibinfo {author} {\bibfnamefont {J.~A.}\ \bibnamefont {Mol}},
  \bibinfo {author} {\bibfnamefont {R.}~\bibnamefont {Rahman}}, \bibinfo
  {author} {\bibfnamefont {G.}~\bibnamefont {Klimeck}}, \bibinfo {author}
  {\bibfnamefont {M.~Y.}\ \bibnamefont {Simmons}}, \bibinfo {author}
  {\bibfnamefont {L.~C.~L.}\ \bibnamefont {Hollenberg}},\ and\ \bibinfo
  {author} {\bibfnamefont {S.}~\bibnamefont {Rogge}},\ }\href
  {https://doi.org/10.1038/ncomms11342} {\bibfield  {journal} {\bibinfo
  {journal} {Nature Communications}\ }\textbf {\bibinfo {volume} {7}},\
  \bibinfo {pages} {11342} (\bibinfo {year} {2016})}\BibitemShut {NoStop}%
\bibitem [{\citenamefont {Hensgens}\ \emph {et~al.}(2017)\citenamefont
  {Hensgens}, \citenamefont {Fujita}, \citenamefont {Janssen}, \citenamefont
  {Li}, \citenamefont {Van~Diepen}, \citenamefont {Reichl}, \citenamefont
  {Wegscheider}, \citenamefont {Das~Sarma},\ and\ \citenamefont
  {Vandersypen}}]{Hensgens2017}%
  \BibitemOpen
  \bibfield  {author} {\bibinfo {author} {\bibfnamefont {T.}~\bibnamefont
  {Hensgens}}, \bibinfo {author} {\bibfnamefont {T.}~\bibnamefont {Fujita}},
  \bibinfo {author} {\bibfnamefont {L.}~\bibnamefont {Janssen}}, \bibinfo
  {author} {\bibfnamefont {X.}~\bibnamefont {Li}}, \bibinfo {author}
  {\bibfnamefont {C.~J.}\ \bibnamefont {Van~Diepen}}, \bibinfo {author}
  {\bibfnamefont {C.}~\bibnamefont {Reichl}}, \bibinfo {author} {\bibfnamefont
  {W.}~\bibnamefont {Wegscheider}}, \bibinfo {author} {\bibfnamefont
  {S.}~\bibnamefont {Das~Sarma}},\ and\ \bibinfo {author} {\bibfnamefont
  {L.~M.~K.}\ \bibnamefont {Vandersypen}},\ }\href
  {https://doi.org/10.1038/nature23022} {\bibfield  {journal} {\bibinfo
  {journal} {Nature}\ }\textbf {\bibinfo {volume} {548}},\ \bibinfo {pages}
  {70–73} (\bibinfo {year} {2017})}\BibitemShut {NoStop}%
\bibitem [{\citenamefont {Singha}\ \emph {et~al.}(2011)\citenamefont {Singha},
  \citenamefont {Gibertini}, \citenamefont {Karmakar}, \citenamefont {Yuan},
  \citenamefont {Polini}, \citenamefont {Vignale}, \citenamefont {Katsnelson},
  \citenamefont {Pinczuk}, \citenamefont {Pfeiffer}, \citenamefont {West},\
  and\ \citenamefont {et~al.}}]{Singha2011}%
  \BibitemOpen
  \bibfield  {author} {\bibinfo {author} {\bibfnamefont {A.}~\bibnamefont
  {Singha}}, \bibinfo {author} {\bibfnamefont {M.}~\bibnamefont {Gibertini}},
  \bibinfo {author} {\bibfnamefont {B.}~\bibnamefont {Karmakar}}, \bibinfo
  {author} {\bibfnamefont {S.}~\bibnamefont {Yuan}}, \bibinfo {author}
  {\bibfnamefont {M.}~\bibnamefont {Polini}}, \bibinfo {author} {\bibfnamefont
  {G.}~\bibnamefont {Vignale}}, \bibinfo {author} {\bibfnamefont {M.~I.}\
  \bibnamefont {Katsnelson}}, \bibinfo {author} {\bibfnamefont
  {A.}~\bibnamefont {Pinczuk}}, \bibinfo {author} {\bibfnamefont {L.~N.}\
  \bibnamefont {Pfeiffer}}, \bibinfo {author} {\bibfnamefont {K.~W.}\
  \bibnamefont {West}},\ and\ \bibinfo {author} {\bibnamefont {et~al.}},\
  }\href {https://doi.org/10.1126/science.1204333} {\bibfield  {journal}
  {\bibinfo  {journal} {Science}\ }\textbf {\bibinfo {volume} {332}},\ \bibinfo
  {pages} {1176–1179} (\bibinfo {year} {2011})}\BibitemShut {NoStop}%
\bibitem [{\citenamefont {Smith}\ \emph {et~al.}(2019)\citenamefont {Smith},
  \citenamefont {Kim}, \citenamefont {Pollmann},\ and\ \citenamefont
  {Knolle}}]{smith2019}%
  \BibitemOpen
  \bibfield  {author} {\bibinfo {author} {\bibfnamefont {A.}~\bibnamefont
  {Smith}}, \bibinfo {author} {\bibfnamefont {M.~S.}\ \bibnamefont {Kim}},
  \bibinfo {author} {\bibfnamefont {F.}~\bibnamefont {Pollmann}},\ and\
  \bibinfo {author} {\bibfnamefont {J.}~\bibnamefont {Knolle}},\ }\href
  {https://doi.org/10.1038/s41534-019-0217-0} {\bibfield  {journal} {\bibinfo
  {journal} {npj Quantum Information}\ }\textbf {\bibinfo {volume} {5}},\
  \bibinfo {pages} {106} (\bibinfo {year} {2019})}\BibitemShut {NoStop}%
\bibitem [{\citenamefont {Bakr}\ \emph {et~al.}(2009)\citenamefont {Bakr},
  \citenamefont {Gillen}, \citenamefont {Peng}, \citenamefont {Fölling},\ and\
  \citenamefont {Greiner}}]{Bakr2009}%
  \BibitemOpen
  \bibfield  {author} {\bibinfo {author} {\bibfnamefont {W.~S.}\ \bibnamefont
  {Bakr}}, \bibinfo {author} {\bibfnamefont {J.~I.}\ \bibnamefont {Gillen}},
  \bibinfo {author} {\bibfnamefont {A.}~\bibnamefont {Peng}}, \bibinfo {author}
  {\bibfnamefont {S.}~\bibnamefont {Fölling}},\ and\ \bibinfo {author}
  {\bibfnamefont {M.}~\bibnamefont {Greiner}},\ }\href
  {https://doi.org/10.1038/nature08482} {\bibfield  {journal} {\bibinfo
  {journal} {Nature}\ }\textbf {\bibinfo {volume} {462}},\ \bibinfo {pages}
  {74–77} (\bibinfo {year} {2009})}\BibitemShut {NoStop}%
\bibitem [{\citenamefont {Schreiber}\ \emph {et~al.}(2015)\citenamefont
  {Schreiber}, \citenamefont {Hodgman}, \citenamefont {Bordia}, \citenamefont
  {Luschen}, \citenamefont {Fischer}, \citenamefont {Vosk}, \citenamefont
  {Altman}, \citenamefont {Schneider},\ and\ \citenamefont
  {Bloch}}]{Schreiber2015}%
  \BibitemOpen
  \bibfield  {author} {\bibinfo {author} {\bibfnamefont {M.}~\bibnamefont
  {Schreiber}}, \bibinfo {author} {\bibfnamefont {S.~S.}\ \bibnamefont
  {Hodgman}}, \bibinfo {author} {\bibfnamefont {P.}~\bibnamefont {Bordia}},
  \bibinfo {author} {\bibfnamefont {H.~P.}\ \bibnamefont {Luschen}}, \bibinfo
  {author} {\bibfnamefont {M.~H.}\ \bibnamefont {Fischer}}, \bibinfo {author}
  {\bibfnamefont {R.}~\bibnamefont {Vosk}}, \bibinfo {author} {\bibfnamefont
  {E.}~\bibnamefont {Altman}}, \bibinfo {author} {\bibfnamefont
  {U.}~\bibnamefont {Schneider}},\ and\ \bibinfo {author} {\bibfnamefont
  {I.}~\bibnamefont {Bloch}},\ }\href {https://doi.org/10.1126/science.aaa7432}
  {\bibfield  {journal} {\bibinfo  {journal} {Science}\ }\textbf {\bibinfo
  {volume} {349}},\ \bibinfo {pages} {842–845} (\bibinfo {year}
  {2015})}\BibitemShut {NoStop}%
\bibitem [{\citenamefont {Choi}\ \emph {et~al.}(2016)\citenamefont {Choi},
  \citenamefont {Hild}, \citenamefont {Zeiher}, \citenamefont {Schauss},
  \citenamefont {Rubio-Abadal}, \citenamefont {Yefsah}, \citenamefont
  {Khemani}, \citenamefont {Huse}, \citenamefont {Bloch},\ and\ \citenamefont
  {Gross}}]{Choi2016}%
  \BibitemOpen
  \bibfield  {author} {\bibinfo {author} {\bibfnamefont {J.-y.}\ \bibnamefont
  {Choi}}, \bibinfo {author} {\bibfnamefont {S.}~\bibnamefont {Hild}}, \bibinfo
  {author} {\bibfnamefont {J.}~\bibnamefont {Zeiher}}, \bibinfo {author}
  {\bibfnamefont {P.}~\bibnamefont {Schauss}}, \bibinfo {author} {\bibfnamefont
  {A.}~\bibnamefont {Rubio-Abadal}}, \bibinfo {author} {\bibfnamefont
  {T.}~\bibnamefont {Yefsah}}, \bibinfo {author} {\bibfnamefont
  {V.}~\bibnamefont {Khemani}}, \bibinfo {author} {\bibfnamefont {D.~A.}\
  \bibnamefont {Huse}}, \bibinfo {author} {\bibfnamefont {I.}~\bibnamefont
  {Bloch}},\ and\ \bibinfo {author} {\bibfnamefont {C.}~\bibnamefont {Gross}},\
  }\href {https://doi.org/10.1126/science.aaf8834} {\bibfield  {journal}
  {\bibinfo  {journal} {Science}\ }\textbf {\bibinfo {volume} {352}},\ \bibinfo
  {pages} {1547–1552} (\bibinfo {year} {2016})}\BibitemShut {NoStop}%
\bibitem [{\citenamefont {Bordia}\ \emph {et~al.}(2017)\citenamefont {Bordia},
  \citenamefont {Lüschen}, \citenamefont {Scherg}, \citenamefont
  {Gopalakrishnan}, \citenamefont {Knap}, \citenamefont {Schneider},\ and\
  \citenamefont {Bloch}}]{Bordia2017}%
  \BibitemOpen
  \bibfield  {author} {\bibinfo {author} {\bibfnamefont {P.}~\bibnamefont
  {Bordia}}, \bibinfo {author} {\bibfnamefont {H.}~\bibnamefont {Lüschen}},
  \bibinfo {author} {\bibfnamefont {S.}~\bibnamefont {Scherg}}, \bibinfo
  {author} {\bibfnamefont {S.}~\bibnamefont {Gopalakrishnan}}, \bibinfo
  {author} {\bibfnamefont {M.}~\bibnamefont {Knap}}, \bibinfo {author}
  {\bibfnamefont {U.}~\bibnamefont {Schneider}},\ and\ \bibinfo {author}
  {\bibfnamefont {I.}~\bibnamefont {Bloch}},\ }\href
  {https://doi.org/10.1103/physrevx.7.041047} {\bibfield  {journal} {\bibinfo
  {journal} {Physical Review X}\ }\textbf {\bibinfo {volume} {7}},\ \bibinfo
  {pages} {041047} (\bibinfo {year} {2017})}\BibitemShut {NoStop}%
\bibitem [{\citenamefont {Mitra}\ \emph {et~al.}(2017)\citenamefont {Mitra},
  \citenamefont {Brown}, \citenamefont {Guardado-Sanchez}, \citenamefont
  {Kondov}, \citenamefont {Devakul}, \citenamefont {Huse}, \citenamefont
  {Schauß},\ and\ \citenamefont {Bakr}}]{Mitra2017}%
  \BibitemOpen
  \bibfield  {author} {\bibinfo {author} {\bibfnamefont {D.}~\bibnamefont
  {Mitra}}, \bibinfo {author} {\bibfnamefont {P.~T.}\ \bibnamefont {Brown}},
  \bibinfo {author} {\bibfnamefont {E.}~\bibnamefont {Guardado-Sanchez}},
  \bibinfo {author} {\bibfnamefont {S.~S.}\ \bibnamefont {Kondov}}, \bibinfo
  {author} {\bibfnamefont {T.}~\bibnamefont {Devakul}}, \bibinfo {author}
  {\bibfnamefont {D.~A.}\ \bibnamefont {Huse}}, \bibinfo {author}
  {\bibfnamefont {P.}~\bibnamefont {Schauß}},\ and\ \bibinfo {author}
  {\bibfnamefont {W.~S.}\ \bibnamefont {Bakr}},\ }\href
  {https://doi.org/10.1038/nphys4297} {\bibfield  {journal} {\bibinfo
  {journal} {Nature Physics}\ }\textbf {\bibinfo {volume} {14}},\ \bibinfo
  {pages} {173–177} (\bibinfo {year} {2017})}\BibitemShut {NoStop}%
\bibitem [{\citenamefont {Peruzzo}\ \emph {et~al.}(2014)\citenamefont
  {Peruzzo}, \citenamefont {McClean}, \citenamefont {Shadbolt}, \citenamefont
  {Yung}, \citenamefont {Zhou}, \citenamefont {Love}, \citenamefont
  {Aspuru-Guzik},\ and\ \citenamefont {O’Brien}}]{Peruzzo2014}%
  \BibitemOpen
  \bibfield  {author} {\bibinfo {author} {\bibfnamefont {A.}~\bibnamefont
  {Peruzzo}}, \bibinfo {author} {\bibfnamefont {J.}~\bibnamefont {McClean}},
  \bibinfo {author} {\bibfnamefont {P.}~\bibnamefont {Shadbolt}}, \bibinfo
  {author} {\bibfnamefont {M.-H.}\ \bibnamefont {Yung}}, \bibinfo {author}
  {\bibfnamefont {X.-Q.}\ \bibnamefont {Zhou}}, \bibinfo {author}
  {\bibfnamefont {P.~J.}\ \bibnamefont {Love}}, \bibinfo {author}
  {\bibfnamefont {A.}~\bibnamefont {Aspuru-Guzik}},\ and\ \bibinfo {author}
  {\bibfnamefont {J.~L.}\ \bibnamefont {O’Brien}},\ }\href
  {https://doi.org/10.1038/ncomms5213} {\bibfield  {journal} {\bibinfo
  {journal} {Nature Communications}\ }\textbf {\bibinfo {volume} {5}},\
  \bibinfo {pages} {4213} (\bibinfo {year} {2014})}\BibitemShut {NoStop}%
\bibitem [{\citenamefont {McClean}\ \emph {et~al.}(2016)\citenamefont
  {McClean}, \citenamefont {Romero}, \citenamefont {Babbush},\ and\
  \citenamefont {Aspuru-Guzik}}]{McClean2016}%
  \BibitemOpen
  \bibfield  {author} {\bibinfo {author} {\bibfnamefont {J.~R.}\ \bibnamefont
  {McClean}}, \bibinfo {author} {\bibfnamefont {J.}~\bibnamefont {Romero}},
  \bibinfo {author} {\bibfnamefont {R.}~\bibnamefont {Babbush}},\ and\ \bibinfo
  {author} {\bibfnamefont {A.}~\bibnamefont {Aspuru-Guzik}},\ }\href
  {https://doi.org/10.1088/1367-2630/18/2/023023} {\bibfield  {journal}
  {\bibinfo  {journal} {New Journal of Physics}\ }\textbf {\bibinfo {volume}
  {18}},\ \bibinfo {pages} {023023} (\bibinfo {year} {2016})}\BibitemShut
  {NoStop}%
\bibitem [{\citenamefont {Kandala}\ \emph {et~al.}(2017)\citenamefont
  {Kandala}, \citenamefont {Mezzacapo}, \citenamefont {Temme}, \citenamefont
  {Takita}, \citenamefont {Brink}, \citenamefont {Chow},\ and\ \citenamefont
  {Gambetta}}]{Kandala2017}%
  \BibitemOpen
  \bibfield  {author} {\bibinfo {author} {\bibfnamefont {A.}~\bibnamefont
  {Kandala}}, \bibinfo {author} {\bibfnamefont {A.}~\bibnamefont {Mezzacapo}},
  \bibinfo {author} {\bibfnamefont {K.}~\bibnamefont {Temme}}, \bibinfo
  {author} {\bibfnamefont {M.}~\bibnamefont {Takita}}, \bibinfo {author}
  {\bibfnamefont {M.}~\bibnamefont {Brink}}, \bibinfo {author} {\bibfnamefont
  {J.~M.}\ \bibnamefont {Chow}},\ and\ \bibinfo {author} {\bibfnamefont
  {J.~M.}\ \bibnamefont {Gambetta}},\ }\href
  {https://doi.org/10.1038/nature23879} {\bibfield  {journal} {\bibinfo
  {journal} {Nature}\ }\textbf {\bibinfo {volume} {549}},\ \bibinfo {pages}
  {242–246} (\bibinfo {year} {2017})}\BibitemShut {NoStop}%
\bibitem [{\citenamefont {Yuan}\ \emph {et~al.}(2021)\citenamefont {Yuan},
  \citenamefont {Sun}, \citenamefont {Liu}, \citenamefont {Zhao},\ and\
  \citenamefont {Zhou}}]{Yuan2021}%
  \BibitemOpen
  \bibfield  {author} {\bibinfo {author} {\bibfnamefont {X.}~\bibnamefont
  {Yuan}}, \bibinfo {author} {\bibfnamefont {J.}~\bibnamefont {Sun}}, \bibinfo
  {author} {\bibfnamefont {J.}~\bibnamefont {Liu}}, \bibinfo {author}
  {\bibfnamefont {Q.}~\bibnamefont {Zhao}},\ and\ \bibinfo {author}
  {\bibfnamefont {Y.}~\bibnamefont {Zhou}},\ }\bibfield  {journal} {\bibinfo
  {journal} {Physical Review Letters}\ }\textbf {\bibinfo {volume} {127}},\
  \href {https://doi.org/10.1103/physrevlett.127.040501}
  {10.1103/physrevlett.127.040501} (\bibinfo {year} {2021})\BibitemShut
  {NoStop}%
\bibitem [{\citenamefont {Peng}\ \emph {et~al.}(2020)\citenamefont {Peng},
  \citenamefont {Harrow}, \citenamefont {Ozols},\ and\ \citenamefont
  {Wu}}]{Peng2020}%
  \BibitemOpen
  \bibfield  {author} {\bibinfo {author} {\bibfnamefont {T.}~\bibnamefont
  {Peng}}, \bibinfo {author} {\bibfnamefont {A.~W.}\ \bibnamefont {Harrow}},
  \bibinfo {author} {\bibfnamefont {M.}~\bibnamefont {Ozols}},\ and\ \bibinfo
  {author} {\bibfnamefont {X.}~\bibnamefont {Wu}},\ }\bibfield  {journal}
  {\bibinfo  {journal} {Physical Review Letters}\ }\textbf {\bibinfo {volume}
  {125}},\ \href {https://doi.org/10.1103/physrevlett.125.150504}
  {10.1103/physrevlett.125.150504} (\bibinfo {year} {2020})\BibitemShut
  {NoStop}%
\bibitem [{\citenamefont {Huang}\ \emph {et~al.}(2020)\citenamefont {Huang},
  \citenamefont {Kueng},\ and\ \citenamefont {Preskill}}]{Haung2020}%
  \BibitemOpen
  \bibfield  {author} {\bibinfo {author} {\bibfnamefont {H.-Y.}\ \bibnamefont
  {Huang}}, \bibinfo {author} {\bibfnamefont {R.}~\bibnamefont {Kueng}},\ and\
  \bibinfo {author} {\bibfnamefont {J.}~\bibnamefont {Preskill}},\ }\href
  {https://doi.org/10.1038/s41567-020-0932-7} {\bibfield  {journal} {\bibinfo
  {journal} {Nature Physics}\ }\textbf {\bibinfo {volume} {16}},\ \bibinfo
  {pages} {1050–1057} (\bibinfo {year} {2020})}\BibitemShut {NoStop}%
\bibitem [{\citenamefont {Paini}\ \emph {et~al.}(2021)\citenamefont {Paini},
  \citenamefont {Kalev}, \citenamefont {Padilha},\ and\ \citenamefont
  {Ruck}}]{Paini2021}%
  \BibitemOpen
  \bibfield  {author} {\bibinfo {author} {\bibfnamefont {M.}~\bibnamefont
  {Paini}}, \bibinfo {author} {\bibfnamefont {A.}~\bibnamefont {Kalev}},
  \bibinfo {author} {\bibfnamefont {D.}~\bibnamefont {Padilha}},\ and\ \bibinfo
  {author} {\bibfnamefont {B.}~\bibnamefont {Ruck}},\ }\href
  {https://doi.org/10.22331/q-2021-03-16-413} {\bibfield  {journal} {\bibinfo
  {journal} {Quantum}\ }\textbf {\bibinfo {volume} {5}},\ \bibinfo {pages}
  {413} (\bibinfo {year} {2021})}\BibitemShut {NoStop}%
\bibitem [{\citenamefont {Fischer}\ and\ \citenamefont
  {Gunlycke}(2019)}]{Fischer2019}%
  \BibitemOpen
  \bibfield  {author} {\bibinfo {author} {\bibfnamefont {S.~A.}\ \bibnamefont
  {Fischer}}\ and\ \bibinfo {author} {\bibfnamefont {D.}~\bibnamefont
  {Gunlycke}},\ }\href@noop {} {\  (\bibinfo {year} {2019})},\ \Eprint
  {https://arxiv.org/abs/1907.01493} {arXiv:1907.01493 [quant-ph]} \BibitemShut
  {NoStop}%
\bibitem [{\citenamefont {Stenger}\ \emph {et~al.}(2021)\citenamefont
  {Stenger}, \citenamefont {Bronn}, \citenamefont {Egger},\ and\ \citenamefont
  {Pekker}}]{stenger2021}%
  \BibitemOpen
  \bibfield  {author} {\bibinfo {author} {\bibfnamefont {J.~P.~T.}\
  \bibnamefont {Stenger}}, \bibinfo {author} {\bibfnamefont {N.~T.}\
  \bibnamefont {Bronn}}, \bibinfo {author} {\bibfnamefont {D.~J.}\ \bibnamefont
  {Egger}},\ and\ \bibinfo {author} {\bibfnamefont {D.}~\bibnamefont
  {Pekker}},\ }\href {https://doi.org/10.1103/PhysRevResearch.3.033171}
  {\bibfield  {journal} {\bibinfo  {journal} {Phys. Rev. Research}\ }\textbf
  {\bibinfo {volume} {3}},\ \bibinfo {pages} {033171} (\bibinfo {year}
  {2021})}\BibitemShut {NoStop}%
\bibitem [{\citenamefont {Bravyi}\ \emph {et~al.}(2017)\citenamefont {Bravyi},
  \citenamefont {Gambetta}, \citenamefont {Mezzacapo},\ and\ \citenamefont
  {Temme}}]{bravyi2017}%
  \BibitemOpen
  \bibfield  {author} {\bibinfo {author} {\bibfnamefont {S.}~\bibnamefont
  {Bravyi}}, \bibinfo {author} {\bibfnamefont {J.~M.}\ \bibnamefont
  {Gambetta}}, \bibinfo {author} {\bibfnamefont {A.}~\bibnamefont
  {Mezzacapo}},\ and\ \bibinfo {author} {\bibfnamefont {K.}~\bibnamefont
  {Temme}},\ }\href@noop {} {\  (\bibinfo {year} {2017})},\ \Eprint
  {https://arxiv.org/abs/1701.08213} {arXiv:1701.08213 [quant-ph]} \BibitemShut
  {NoStop}%
\bibitem [{\citenamefont {{Jordan}}\ and\ \citenamefont
  {{Wigner}}(1928)}]{jordan1928}%
  \BibitemOpen
  \bibfield  {author} {\bibinfo {author} {\bibfnamefont {P.}~\bibnamefont
  {{Jordan}}}\ and\ \bibinfo {author} {\bibfnamefont {E.}~\bibnamefont
  {{Wigner}}},\ }\href {https://doi.org/10.1007/BF01331938} {\bibfield
  {journal} {\bibinfo  {journal} {Zeitschrift fur Physik}\ }\textbf {\bibinfo
  {volume} {47}},\ \bibinfo {pages} {631} (\bibinfo {year} {1928})}\BibitemShut
  {NoStop}%
\bibitem [{\citenamefont {Seeley}\ \emph {et~al.}(2012)\citenamefont {Seeley},
  \citenamefont {Richard},\ and\ \citenamefont {Love}}]{Seeley2012}%
  \BibitemOpen
  \bibfield  {author} {\bibinfo {author} {\bibfnamefont {J.~T.}\ \bibnamefont
  {Seeley}}, \bibinfo {author} {\bibfnamefont {M.~J.}\ \bibnamefont
  {Richard}},\ and\ \bibinfo {author} {\bibfnamefont {P.~J.}\ \bibnamefont
  {Love}},\ }\href {https://doi.org/10.1063/1.4768229} {\bibfield  {journal}
  {\bibinfo  {journal} {The Journal of Chemical Physics}\ }\textbf {\bibinfo
  {volume} {137}},\ \bibinfo {pages} {224109} (\bibinfo {year}
  {2012})}\BibitemShut {NoStop}%
\bibitem [{\citenamefont {Bravyi}\ and\ \citenamefont
  {Kitaev}(2002)}]{Bravyi2002}%
  \BibitemOpen
  \bibfield  {author} {\bibinfo {author} {\bibfnamefont {S.~B.}\ \bibnamefont
  {Bravyi}}\ and\ \bibinfo {author} {\bibfnamefont {A.~Y.}\ \bibnamefont
  {Kitaev}},\ }\href {https://doi.org/10.1006/aphy.2002.6254} {\bibfield
  {journal} {\bibinfo  {journal} {Annals of Physics}\ }\textbf {\bibinfo
  {volume} {298}},\ \bibinfo {pages} {210–226} (\bibinfo {year}
  {2002})}\BibitemShut {NoStop}%
\bibitem [{\citenamefont {Lin}\ \emph {et~al.}(1993)\citenamefont {Lin},
  \citenamefont {Gubernatis}, \citenamefont {Gould},\ and\ \citenamefont
  {Tobochnik}}]{Lin1993}%
  \BibitemOpen
  \bibfield  {author} {\bibinfo {author} {\bibfnamefont {H.}~\bibnamefont
  {Lin}}, \bibinfo {author} {\bibfnamefont {J.}~\bibnamefont {Gubernatis}},
  \bibinfo {author} {\bibfnamefont {H.}~\bibnamefont {Gould}},\ and\ \bibinfo
  {author} {\bibfnamefont {J.}~\bibnamefont {Tobochnik}},\ }\href
  {https://doi.org/10.1063/1.4823192} {\bibfield  {journal} {\bibinfo
  {journal} {Computers in Physics}\ }\textbf {\bibinfo {volume} {7}},\ \bibinfo
  {pages} {400} (\bibinfo {year} {1993})}\BibitemShut {NoStop}%
\bibitem [{\citenamefont {Gunlycke}\ \emph {et~al.}(2020)\citenamefont
  {Gunlycke}, \citenamefont {Palenik}, \citenamefont {Emmert},\ and\
  \citenamefont {Fischer}}]{gunlycke2020}%
  \BibitemOpen
  \bibfield  {author} {\bibinfo {author} {\bibfnamefont {D.}~\bibnamefont
  {Gunlycke}}, \bibinfo {author} {\bibfnamefont {M.~C.}\ \bibnamefont
  {Palenik}}, \bibinfo {author} {\bibfnamefont {A.~R.}\ \bibnamefont
  {Emmert}},\ and\ \bibinfo {author} {\bibfnamefont {S.~A.}\ \bibnamefont
  {Fischer}},\ }\href@noop {} {\  (\bibinfo {year} {2020})},\ \Eprint
  {https://arxiv.org/abs/2011.08942} {arXiv:2011.08942 [quant-ph]} \BibitemShut
  {NoStop}%
\end{thebibliography}%
\bibliographystyle{apsrev4-2}

\end{document}